# A REMOVABLE TEMPERATURE SENSORS HOLSDER FOR CRYOGENIC RAYLEIGH-BÉNARD CONVECTION CELL

**Pavel HANZELKA, Ivan VLCEK, Pavel URBAN**

The Czech Academy of Sciences, Institute of Scientific Instruments
Brno, 612 00, Czech Republic, urban@isibrno.cz

## ABSTRACT

A removable holder carrying a set of miniature temperature sensors inside our Rayleigh-Bénard convection cylindrical cell has been developed. The sensors should not disturb the temperature and velocity fields of the buoyancy-driven turbulent flow of our working fluid - cryogenic helium gas (~ 5 K). The cell, 300 mm both in diameter and height, must withstand the overpressure up to 2.5 bar, which is why all its parts are welded together. The holder consists of a pair of identical feedthroughs (each with 6 glass-sealed pins), constantan wires (0.1 mm), Be bronze springs and insulating elements and enables installing up to 6 sensors simultaneously, with the option to replace the damaged ones. We already used two such holders and obtained a map of local temperature and its fluctuations in 12 positions inside the cell.

## 1. INTRODUCTION

Turbulent convection of fluids induced by temperature gradient is a common phenomenon that occurs in many natural systems. Nevertheless, a general theory fully elucidating turbulent convection is still not completed. Experimental studies enable us to enumerate dimensionless numbers characterizing the convection and compare the results with theory. The simplest model, so called Rayleigh-Bénard convection (RBC), is usually used (Tritton 1988). RBC is defined as convection of a fluid enclosed between two infinite isothermal horizontal plates placed in a gravitational field. The convection is induced by heating of the lower plate and cooling of the upper one (Ahlers at al. 2009, Chilla and Schumacher 2012, Xia 2013).

If the turbulent convection of large natural systems, e.g. in atmosphere or in oceans should be experimentally simulated in a laboratory, the values of the characteristic dimensionless numbers of the model must be comparable with those of the modelled system (Chilla and Schumacher 2012). One of the dimensionless numbers concerning the RBC, the Rayleigh number, Ra, depends on the third power of the vertical distance $L$ between the plates:

$$Ra = \frac{\alpha \cdot g \cdot \Delta T \cdot L^3}{\nu \cdot \kappa} \qquad Equation\ (1)$$

Here $g$ is the gravity acceleration, $\Delta T$ is the temperature difference between the plates. The properties of the fluid are characterized by the isobaric thermal expansion coefficient, $\alpha$, kinematic viscosity, $\nu$, and thermal diffusivity, $\kappa$ (Ahlers at al. 2009, Chilla and Schumacher 2012). Thus, if too big spatial dimensions of the experimental model could cause problems, the needed value of $Ra$ can be reached by using of a fluid with appropriate values of $\alpha$, $\nu$, and $\kappa$. Cryogenic helium in state of gas or supercritical fluid (critical point: 5.19 K, 227 kPa) is suitable for those experiments.

Our group conducts experiments with cryogenic helium at temperatures 4.2 K to 6 K using a cylindrical RBC cell of 300 mm in diameter and 150 mm to 300 mm in height designed for pressure up to 250 kPa (Urban et al. 2010). Later a set of miniature temperature sensors (Mitin et al. 2007) placed inside of the cell was used for local measurement of temperatures and its fluctuations in turbulent flow of cryogenic helium (Urban et al.

2014, Skrbek and Urban 2015, Musilova et al. 2017). To enable easy installation and replacement of the sensors, the upper and lower plate were designed as demountable with indium-sealed flanges. However, at higher pressures, leaks of the seals occurred in the course of repeated experiments. Therefore, we re-designed the RBC cell as shown in Fig. 1. The flanges of the thin sidewalls attached to the plates and the flanges of the central part of the cell are now welded together. In consequence, the necessity to eliminate the destruction of the overvoltage sensitive sensors during welding prompted us to design a removable sensor holder that enables to install a set of sensors into the finished cell and possibly to replace a damaged sensor.

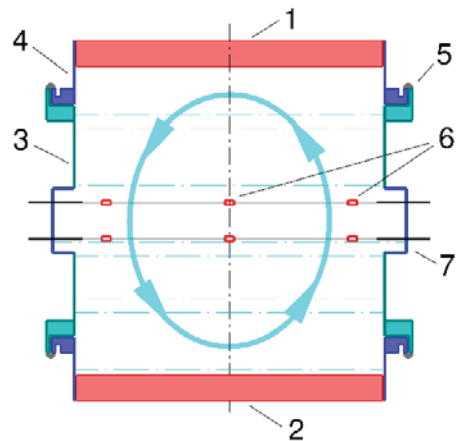

**Figure 1: General sketch of an experimental cryogenic Rayleigh-Bénard convection cell.**

**The cell, placed in high vacuum and surrounded by a cold radiation shield (not shown in this figure), consists of two highly conductive copper plates (1, 2) and a cylindrical sidewall (3). To reduce the parasitic heat flows, the collars (4) adjacent to the plates are made as thin as possible. All the main parts are welded together (5). The convection is forced by heating of the lower plate (2) and cooling of the upper one (1). Miniature temperature sensors (6) connected to vacuum feedthroughs (7) are used for measurement of the temperature and its fluctuations in the convecting medium**

## 2. SENSORS HOLDER DEVELOPEMENT

### 2.1. Sensors holder requirements

The sensor holder ought to fulfil the following requirements:

- Each complete holder should support up to 6 sensors.

- Individual sensors are to be placed in two levels, both in the centre and near to the sidewall of the cell.

- The presence of the sensors and their current leads should not significantly affect the temperature and velocity fields of the convecting medium.

- Repeated installation and removal of the holder into the welded cell is required.

### 2.2. Holder design

Fig. 2 shows the main parts of the holder including the sensors while Fig. 3 represents a more detailed drawing of the cryogenic vacuum feedthrough.

In general, two identical vacuum feedthroughs, each of 6 pins, are attached to the sidewall of the RBC cell. Four constantan wires of 0.1 mm in diameter spanned between the pins of opposite feedthroughs support the sensors as shown in Fig. 2. Each of the wires is divided into three sections by miniature electrically insulating elements. Additional wires joined to the middle wire sections connect the central sensors with the fifth and the sixth pins of each feedthrough (Fig. 3). Thanks to the low thermal conductivity of constantan, the wires do not affect the temperature field of the convecting fluid. On the other hand, the electrical resistance of the wires is high and the temperature sensors must be calibrated in situ, including their leads.

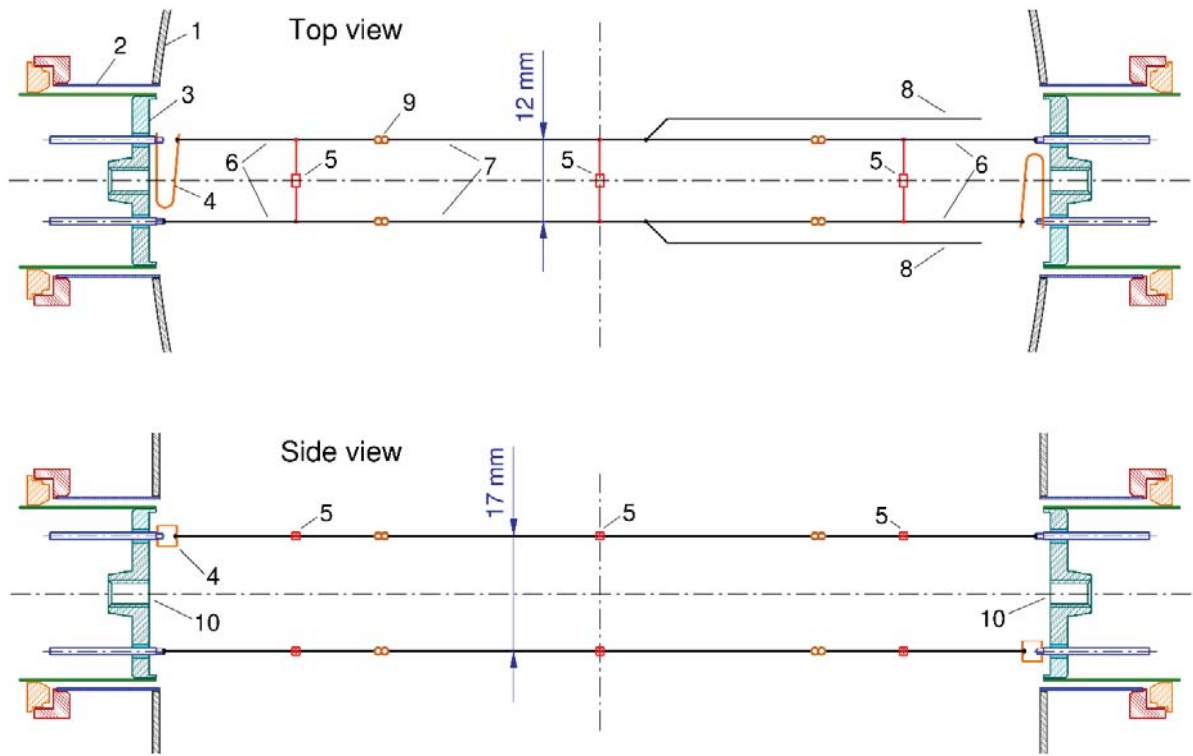

**Figure 2: Arrangement of the temperature sensors holder.**

**1 - sidewall of the cell; 2 - sleeve; 3 - six pins feedthrough; 4 - flat spring; 5 - temperature sensors; 6 - bearing and current leading constantan wires (0.1 mm in diameter) of the side sensors; 7 - bearing wires of the central sensors; 8 - current leading wires of the central sensors, connected to the fifth and the sixth pins of the feedthrough; 9 - miniature electrically insulating elements**

During the holder assembly, both feedthroughs are screwed at a proper distance to an auxiliary threaded rod (central threaded hole M4 in the feedthrough body is used) and then the carrying wires are soldered to the pins and tension springs. After that, the sensor leads are contacted to the wires. The completed holder is then pulled through the sleeves attached to opposite sides of the cylindrical sidewall. As the finishing steps of the holder installation, the feedthroughs are soldered to the sleeves (Fig.3), the auxiliary rod is removed and the threaded holes are plugged with the copper sealing element soldered to a bolt. Reversed procedure is used during the holder removal.

Fig. 3 shows the details of the feedthrough and its connection to the cell sidewall. Six kovar pins in glass-to-metal seals are attached to the feedthrough body made of stainless steel.

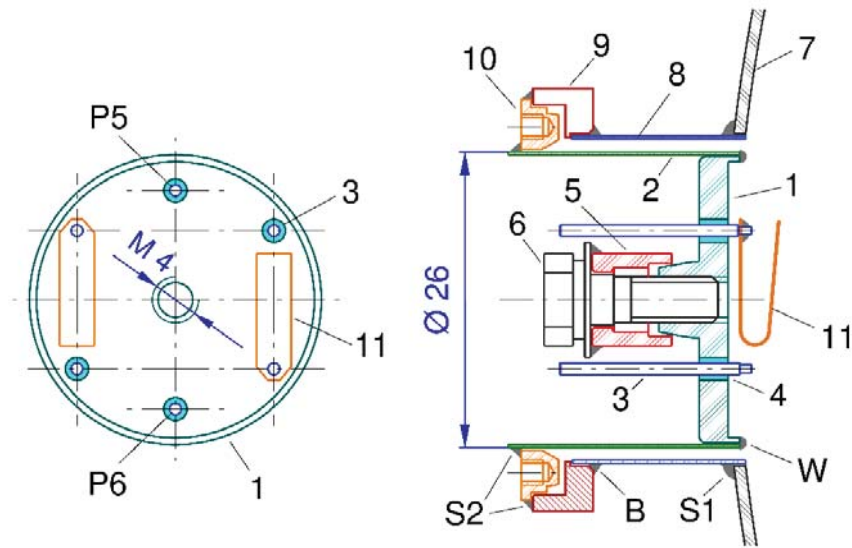

**Figure 3: Detailed sketch of the vacuum feedthrough.**

1 – feedthrough body; 2 – thin wall tube welded (W) to the body; 3 – kovar pin in glass-to-metal seal (4); 5 – copper sealing element soldered to a bolt (6); 7 – cell sidewall; 8 – sleeve with brazed (B) copper ring (9) soldered (S1) to the cell wall; 10 – connecting copper ring with soldered junctions (S2); 11 - beryllium copper flat spring; P5, P6 – pins for connection of the central sensors

Highly polished cone created on the body enables the plugging of the central hole using a copper sealing element. The body with pins is welded to a thin-wall stainless steel tube. This tube protects the glass-to-metal seals against mechanical stress caused by thermal contractions of individual parts during assembly and cooldown. A sleeve consisting of a thin-wall stainless steel tube and a copper ring is soldered to the cell sidewall prior to final holder installation. The connection of the feedthrough to the sleeve provides a copper ring soldered finally between these parts. Threaded holes on its front enable easier removal of the ring during disassembly.

### 2.3. Feedthrough testing and application

The glass to metal seals together with the welded, brazed and soldered junctions must remain leak-tight under the experiment working conditions, i.e. temperature of about 5 K and pressure up to 250 kPa. Therefore, we constructed a device for leak testing of the cooled and pressurised feedthrough. We used cooling with liquid nitrogen instead of liquid helium for simplification. On the other hand, we pressurised each feedthrough several times with helium gas up to 450 kPa. No leaks were found.

We used two pairs of the feedthroughs for the assembly of two temperature sensor holders. Both holders were attached to our RBC cell, one above the other, in perpendicular mutual position. We thus obtained information of the convecting medium temperature and its fluctuations in 12 positions.

### 3. CONCLUSIONS

Each of the developed holder of miniature temperature sensors enables inserting up to six sensors into a closed cryogenic helium RBC cell. The holder may be removed from the cell and the sensors replaced if necessary. The dimensions of the sensors and their suspension are fine enough not to significantly affect the temperature and velocity fields of the convecting medium. We observed no vacuum leaks caused by the holder during long term cryogenic experiments. Valuable data of the temperature field and its fluctuations have been obtained (Urban et al. 2021).

## ACKNOWLEDGEMENTS

We thank T. Králík, M. Macek, V. Musilová and A. Srnka for stimulating discussions. We acknowledge the support of this research by the Czech Science Foundation under the project GAČR GA20-00918S.